**Reanalysis of Uranus' cloud scattering properties from IRTF/SpeX observations using a self-consistent scattering cloud retrieval scheme.**

P.G.J. Irwin[a], D.S. Tice[a], L.N. Fletcher[a], J.K. Barstow[a], N.A. Teanby[b], G.S. Orton[c], and G.R. Davis[d].

[a]Atmospheric, Oceanic, and Planetary Physics, Department of Physics, University of Oxford, Clarendon Laboratory, Parks Road, Oxford, OX1 3PU, United Kingdom.
*Tel:* (+44) 1865 272933, *Fax:* (+44) 1865 272923. E-mail: irwin@atm.ox.ac.uk

[b]School of Earth Sciences, University of Bristol, Wills Memorial Building, Queen's Road, Bristol, BS8 1RJ, United Kingdom.

[c]Jet Propulsion Laboratory, California Institute of Technology, 4800 Oak Grove Drive, Pasadena, CA 91109, USA.

[d]Joint Astronomy Centre, 660 N. A'ohoku Place, Hilo, Hawaii HI 96720, USA.

Original: 17/10/14: Revised 12/12/14

Number of manuscript pages: 28 + acknowledgements, references, figure legends and tables (52 in total).

Number of Figures: 14

Number of Tables: 5

Editorial correspondence should be directed to:

Prof. Patrick G. J. Irwin,

Atmospheric, Oceanic and Planetary Physics, Clarendon Laboratory, Parks Road,

Oxford OX1 3PU, United Kingdom.

*Telephone:*   (+44) 1865 272933   (direct line: 272083)

*Fax:*   (+44) 1865 272923

*Email:*   irwin@atm.ox.ac.uk

**Abstract**

We have developed a new retrieval approach to modelling near-infrared spectra of Uranus that represents a significant improvement over previous modelling methods. We reanalysed IRTF/SpeX observations of Uranus observed in 2009 covering the wavelength range 0.8 – 1.8 μm and reported by Tice et al. (2013). By retrieving the imaginary refractive index spectra of cloud particles we are able to consistently define the real part of the refractive index spectra, through a Kramers-Kronig analysis, and thus determine self-consistent extinction cross-section, single-scattering and phase-function spectra for the clouds and hazes in Uranus' atmosphere. We tested two different cloud-modelling schemes used in conjunction with the temperature/methane profile of Baines et al. (1995), a reanalysis of the Voyager-2 radio-occultation observations performed by Sromovsky, Fry and Tomasko (2011), and a recent determination from Spitzer (Orton et al., 2014). We find that both cloud-modelling schemes represent the observed centre-of-disc spectrum of Uranus well, and both require similar cloud scattering properties of the main cloud residing at ~2 bars. However, a modified version of the Sromovsky, Fry and Tomasko (2011) model, with revised spectral properties of the lowest cloud layer, fits slightly better at shorter wavelengths and is more consistent with the expected vertical position of Uranus' methane cloud.

We find that the bulk of the reflected radiance from Uranus arises from a thick cloud at approximately the 2 bar level, composed of particles that are significantly more absorbing at wavelengths $\lambda > 1.0$ μm than they are at shorter wavelengths $\lambda < 1.0$ μm. This spectral information provides a possible constraint on the identity of the main particle type, although we find that the scattering properties required are not consistent with any of the available laboratory data for pure $NH_3$, $NH_4SH$, or $CH_4$ ice (all suspected of condensing in the upper troposphere). It is possible that the observed clouds are mixtures of tropospheric condensate mixed with photochemical products diffusing down from above, which masks their pure scattering features. Because there is no available laboratory data for pure $H_2S$ or $PH_3$ ice (both of which might be present as well), they cannot be excluded as the cloud-forming species. We note, however, that their absorptive properties would have to be two orders of magnitude greater than the other measured ices at wavelengths greater than 1 μm to be consistent with our retrieval, which suggests that mixing with photochemical products may still be important.

## 1. Introduction

The visible/near-infrared spectrum of Uranus is formed by reflection of sunlight from various levels in the planet's atmosphere that are modulated by the transmission of methane gas, $H_2$-$H_2$ (collision-induced absorption) and Rayleigh scattering. It thus provides unique constraints on Uranus' vertical cloud structure and also on the scattering properties of Uranus' hazes and condensates. Many observations of Uranus' near-IR spectrum have been made in the last decade, through Uranus' northern spring equinox in 2007. A number of modelling studies have attempted to interpret these data, including Baines and Bergstrahl (1986) (0.35-1.05 μm), Pollack et al. (1987) (0.43 – 0.6 μm), Rages et al. (1991) (Voyager 2: 0.35 – 0.62 μm), Karkoschka and Tomasko (2009) (HST/STIS: 0.3-1.0 μm), Sromovsky, Irwin and Fry (2006), Sromovsky and Fry (2006, 2008), Sromovsky, Fry and Kim (2011) and Irwin et al. (2007, 2009, 2010, 2011, 2012a,b) (1.2-1.8 μm). One of the primary results of these studies is that the abundance of methane in the upper troposphere of Uranus varies with latitude (Karkoschka and Tomasko, 2009). Since we use methane absorption features to probe the vertical level of the clouds this means that the determined cloud levels are unreliable unless this methane variation is accounted for. Fortunately, this degeneracy can be broken at a few places in Uranus' spectrum, most easily at 0.825 μm (Karkoschka and Tomasko, 2009), where a collision-induced-absorption band of $H_2$-$H_2$ is found. However, the degeneracy can also be broken in the H-band (1.4 – 1.8 μm) if the spectrum is observed at sufficiently high spectral resolution (Irwin et al., 2012b). Most recently, Tice et al. (2013) analysed IRTF/SpeX observations of Uranus made in 2009 and attempted to determine cloud parameters that could be applied over the entire SpeX range of 0.8-1.8 μm using both compact and vertically extended clouds/hazes. Tice et al. (2013) found a variation of methane abundance with latitude that was consistent with Karkoschka and Tomasko (2009), but also found that the particles in the main cloud deck at 2-3 bars must be significantly more absorbing at longer wavelengths with the single-scattering albedo varying from 1.0 for $\lambda < 1.0$ μm to 0.7 for wavelengths longer than 1.4 μm, a conclusion previously reached from UKIRT/UIST observations by Irwin, Teanby and Davis (2010). In this paper we attempt to reconcile the conclusions of Tice et al. (2013) with that of a reanalysis of HST/STIS observations by Sromovsky, Fry and Kim (2011), who approached the problem from the perspective of first determining a vertical profile of methane (and $H_2$, He and Ne) that was consistent with both the Voyager 2 radio-occultation observations (Lindal et al. 1987) and also cloud and haze vertical distributions inferred from observations in the visible and near-IR.

Since the identity of Uranus' cloud particle types is not unambiguously known, previous attempts to model the cloud scattering properties have used empirically adjusted scattering properties, such as assuming extinction cross-section and phase-function spectra, or empirically adjusting the phase function. Although simple, these techniques are rather unphysical and also have usually only been applied to narrow wavelength ranges, making it difficult to match the derived quantities with the laboratory-measured spectra of candidate condensates. In this paper we present a new retrieval scheme that returns a physically based self-consistent determination of the scattering properties of Uranus' clouds. This approach can be applied simultaneously over a wide spectral range and the retrieved cloud parameters can be easily compared with the measured properties of candidate condensates. Section 2 describes our new retrieval scheme and applies it to the IRTF/SpeX observations (0.8-1.8 μm) presented by Tice et al. (2013), using their simple 2-cloud model. We also assess the effects on these retrievals of using the temperature-abundance profiles derived from Spitzer (Orton et al., 2014) and the 'F1' profile of Sromovsky Fry and Kim (2011). Section 3 then compares the observed IRTF/SpeX observations with the predictions of the 5-cloud model and revised temperature-abundance profiles of Sromovsky, Fry and Kim (2011). Section 4 applies our new retrieval scheme to fitting the IRTF/SpeX spectra with a revised model based on the Sromovsky, Fry and Kim (2011) scheme, while section 5 presents a discussion of the derived scattering properties retrieved. Section 6 summarises our conclusions.

**2. Reanalysis of IRTF spectra with a simple two-cloud model**

Tice et al. (2013) (henceforth T2013) analysed SpeX long-slit spectra of Uranus, recorded in 2009 at NASA's Infrared Telescope Facility (IRTF) on Mauna Kea, Hawaii. The slit was aligned with Uranus' central meridian and spectra recorded from 0.8 to 1.8 μm with a spectral resolution of 1200 and with an average 'seeing' that varied from 0.5" in the H-band (1.4-1.8 μm) to 0.6" in the I-band (0.8-0.9 μm). Together with a thin extended haze above the 1-bar pressure level, the main cloud deck in Uranus' atmosphere is found by numerous studies to reside in the 2-3 bar pressure region and is of unknown composition, although from thermochemical models is expected to be formed of $H_2S$ or possibly $NH_3$ ice. Since the precise identification is unknown, and spectroscopic information for condensates such as $H_2S$ is not available, it is necessary to make some assumptions on the particles' scattering properties and different authors have adopted different approaches for how to do this. For example, T2013 assumed that the particles in the main cloud deck and in the haze had the same phase function at all wavelengths (parameterised with a Henyey-Greenstein phase function with an asymmetry parameter g = 0.7) and set the variation of extinction cross-section with wavelength using Mie theory (e.g. Hansen and Travis, 1971) for particles with a complex refractive index of 1.4+0i at all wavelengths and different particle size distributions (a standard Gamma size-distribution was assumed, where $n(r) \propto r^{(1-3b)/b} e^{-r/ab}$, $a$ is the mean radius and $b$ the variance). The single-scattering albedo spectrum was then adjusted empirically at different wavelengths to improve the fit between the measured and synthetic spectra. Other authors, such as Sromovsky, Fry and Kim (2011) (henceforth SFK2011), empirically adjusted the phase function and/or single-scattering albedo as a function of wavelength. We shall return to the approach of SFK2011 later in this paper.

While providing a reasonable starting point for analysing Uranus spectra, the method of T2013, which is based on earlier work by Irwin et al. (2010, 2011, 2012a), is less applicable to data such as IRTF/SpeX that cover a large wavelength range than to small individual wavelength ranges, for which the technique was originally developed. In reality, we expect the phase function to vary with wavelength. Furthermore, it is not very realistic to model the haze, composed of very small particles, with the same phase function as that used for the main cloud deck, composed of micron-sized particles, since small particles would be expected to behave more like Rayleigh scatterers. It is also not very consistent to set the extinction cross-section spectrum equal to that for Mie particles with a single conservative refractive index at all wavelengths and then adjust the single-scattering albedo independently. Thus, in this paper we set out to improve the physical plausibilty of our cloud retrievals by representing the clouds in a more self-consistent manner such that the phase functions used at different wavelengths are consistent with the cross-section and single-scattering albedos. In order to construct this more self-consistent model we needed to define at a more fundamental level the scattering properties of the particles.

For Uranus' clouds we expect there to be some variation of single-scattering albedo, cross-section and phase function with wavelength. However, instead of varying each of these individually, we chose instead to represent this by varying the imaginary part of the particles' complex refractive spectrum. Given the imaginary part of a particle's refractive index spectrum, $n_i$, we can determine the real part, $n_r$, using the Kramers-Kronig analysis (e.g. Sheik-Bahae, 2005), providing we know the value of the real part of the refractive index spectrum at some reference wavelength. We know from laboratory studies that the real part of the refractive index for most particles that may condense in outer planet atmospheres is between 1.3 and 1.4. For example, Martonchik, Orton and Appleby (1984) quote $n_r$ = 1.408 for ammonia ice at 1.46 μm and Martonchik and Orton (1994) quote a value $n_r$ =1.311 for methane ice at the same wavelength. Hence, in this paper we fixed the real part of the refractive index to be nominally 1.4 at an arbitrary reference wavelength, usually 1.4 μm, and determined the value at other wavelengths from the fitted imaginary refractive index spectrum via a Kramers-Kronig analysis. Note that we also tested the effect of assuming other choices of the fixed real refractive index value, which we report later. We would still expect there to be a distribution of particle sizes and so we chose to continue to represent the size distribution with standard Gamma distributions with mean radius and variance. Using the complex refractive index spectrum together with this particle size distribution, standard Mie theory (e.g. Hansen and Travis, 1974) may be used to calculate self-consistent extinction cross-section, single-scattering albedo and phase-function spectra, which can then be used in a radiative transfer model to test against the observed spectra. Hence, by assuming the real part of the refractive index at a single wavelength and fitting the imaginary part of the refractive index spectrum and the parameters of the particle size-distribution, all the other scattering parameters can be calculated completely self-consistently.

We thus added a new parameterisation scheme to our NEMESIS (Irwin et al. 2008) retrieval model, where for each haze or cloud particle type we retrieve the mean radius and variance of the standard Gamma particle-size distribution together with the imaginary refractive index spectrum over the wavelength range of interest. At each iteration NEMESIS takes the latest retrieved values of these parameters, computes the real part of the refractive index spectrum using the Kramers-Kronig analysis and calculates the extinction cross-section, single-scattering albedo and phase-function spectra using Mie theory. NEMESIS then uses these parameters in its existing multiple-scattering model to simulate the observed Uranus IRTF/Spex spectra, although the Mie-scattering calculated phase functions were first approximated with combined Henyey-Greenstein functions to smooth over some features in these phase functions peculiar to purely spherical particles, such as the 'rainbow'. While we would not expect the mean radius and variance to be correlated with any other parameter, we might reasonably expect that the imaginary refractive index spectrum should be a smoothly varying function of wavelength. This smoothing was applied via a correlation length in the *a priori* covariance matrix with off-diagonal elements set as $S_{ij} = (S_{ii}S_{ij})^{1/2} \exp((\lambda_i - \lambda_i)/c)$, where c is a 'correlation length' (which was set to 0.1 μm, to provide wavelength-smoothing on a scale consistent with the laboratory spectra of candidate condensates, shown later in Fig. 14) and $\lambda_i$ is the wavelength of the i[th] refractive index.

While our previous approach of using *ad hoc* cross-section, single-scattering, and Henyey-Greenstein phase-function spectra allowed us great flexibilty in providing plenty of parameters to adjust in order to minimise the closeness of fit between the observed and modelled spectra, this new system has fewer free parameters, but returns a solution that is self-consistent. However, it has to be assumed that the particles can be well modelled with Mie theory. Whilst this is certainly true for spherical droplets, Mie theory is less applicable to solid particle scatterers, such as methane, $H_2S$ and $NH_3$ ice, whose crystals will not be spherical. However, if we assume the ice particles are randomly oriented with respect to each other, then the Mie approach provides a reasonable first-order approximation, although a set of Mie scatterers exibit peaks at angles related to the 'rainbow' and also a higher back-scattering peak (the 'glory'), which are features that are absent from a distribution of non-spherical particles. However, such features are smoothed by our approach of using combined Henyey-Greenstein approximations to the calculated phase functions and our technique of retrieving the imaginary refractive index greatly helps to limit the range of solutions to ones more likely to be physically plausible. We will return to the applicability of the Mie approach in the discussion section and outline possible future refinements there.

To assess the new retrieval scheme we initially adopted the same model as T2013: namely a two-cloud model with the original reference methane profile (Baines et al., 1995) used by Irwin et al. (2007) with a deep volume mixing ratio of 1.6%, limited to 30% relative humidity in the upper troposphere and frozen to its cold trap value in stratosphere. We used the same IRTF/SpeX spectra reported by T2013 and fitted to the spectrum recorded closest to the equator at 1.3°S. The same Matrix-Operator scattering model of Plass et al. (1973) was used, with the inclusion of Rayleigh scattering by the air molecules themselves, with 5 zenith angles and N Fourier components to cover the azimuth variation, where N is set adaptively from the viewing zenith angle, θ, as $N = \text{int}(\theta/3)$. Methane gaseous absorption was modelled with the k-coefficients of Karkoschka and Tomasko (2010) and the $H_2$-$H_2$/$H_2$-He CIA coefficients of Borysow (1991, 1992), Zheng and Borysow (1995) and Borysow et al. (2000) were employed. The contribution of clouds and aerosols was modelled with two discrete clouds, each with a specified base pressure, fractional scale height (relative to the atmospheric pressure scale height) and optical depth. Following on from T2013 the *a priori* main Tropospheric Cloud (TC) was modelled with a base pressure of 2.7 bar, fractional scale height 0.08 and opacity (at 1.6 μm) of 5.5, while the Tropospheric Haze (TH) was modelled with a base pressure of 1 bar, fractional scale height 0.93 and opacity (at 1.6 μm) of 0.027. All six parameters were allowed to vary in these new retrievals with the exception of the haze base pressure, which was fixed at 1 bar, which T2013 found to be the optimal pressure. The tropospheric cloud was assumed to be composed of particles with *a priori* mean radius 1 μm, and variance 0.05, while the haze particles were assumed to be composed of particles with *a priori* mean radius 0.1 μm, with the same variance. The *a priori* imaginary refractive index spectrum of both particles was set to (0.001±0.0005) at all wavelengths, while the real part of the refractive index at 1.4 μm was set to 1.4. For the forward-model calculations, the atmosphere between 17 and 0.01 bar was split into 39 levels of equal thickness in log (pressure) and the cloud parameterisation scheme used to determine the opacity of both cloud types in each layer.

The NEMESIS model was then run until it converged to the solution shown in Fig. 1, where it can be seen that we have achieved a good fit to the IRTF/SpeX spectra across the entire wavelength range (NB the section between 1.35 and 1.46 μm has been omitted due to the low observed signal to noise ratio caused by telluric water absorption). Figure 1 also shows the fit achieved to this spectrum using the original model of T2013 with empirically adjusted scattering properties, together with one with the empirical T2013 properties, but where the assumed temperature and abundance profiles are those determined to best fit Uranus Spitzer observations (Orton et al. 2014), which has a deep methane volume mixing ratio of 3.2%. To achieve these fits, Fig. 2 shows the fitted cloud profiles (optical depth/bar at 1.6 μm), while Fig. 3 shows the retrieved complex refractive index spectra of the Tropospheric Cloud (TC) and Tropospheric Haze (TH) fitted by our model for the self-consistent modelling case, together with the fitted particle size-distribution parameters. These refractive indices are also listed in Table 1. In our model we actually fit values of log($n_i$), in order to ensure that $n_i$ never becomes negative. Our *a priori* values of $n_i$ were 0.001±0.0005, in other words they have a fractional error of 50%, and if we plot log($n_i$), as we do in Fig.3, then retrieved values with the same fractional error have identically large error bars, enabling the reader to see more clearly where the retrieval is giving us meaningful information.

For the main Tropospheric Cloud (TC) deck, we see that we require the particles to become substantially more absorbing at wavelengths greater than 1.0 μm and from the size of the error bars we can see that this requirement is well constrained. At shorter wavelengths we can see that the retrieved values of $n_i$ are very similar to the *a priori*, and the retrieved errors are only slightly smaller, suggesting we have limited sensitivity. For particles in the Tropospheric Haze (TH) we find that our solution barely moves from the *a priori*, indicating that we have little sensitivity to its imaginary refractive index, presumably due to its low optical depth and thus small effect on the measured spectrum.

To assess the sensitivity of our solution to the assumed *a priori* imaginary refractive index, $n_i$, the retrieval was repeated for *a priori* imaginary refractive indices in the range 0.005, 0.01, 0.05, 0.1 (again with *a priori* errors of 50%) for both TC and TH particles. The results of these retrievals (and the $n_i$ = 0.001 case) are shown together in Fig. 4. Considering first the results for the Tropospheric Cloud (TC) particles, we can see that the solutions all converge to increased absorption at wavelengths greater than 1.0 μm. However, the solutions are by no means unique and the spread of results is greater than the formal retrieval errors on individual elements of the solution, which is unsurprising given the non-linearity of this retrieval approach and the various cross-correlations that are absent from the formal errors on individual elements. The mean and standard deviation of all five cases are also shown in Fig. 4 and this represents a more conservative and realistic estimate of the required imaginary refractive index spectrum and its variance. Furthermore, modelling multiple *a priori* $n_i$ values also allows us to determine the likely spread in the real component of the real refractive index. For the Tropospheric Haze (TH) particles we can see that the results do not seem to converge at all, confirming that we are not able to infer meaning information on the TH $n_i$ spectrum using this method. Figure 5 shows the range in extinction cross-section and single-scattering albedo solutions resulting from the 5 different *a priori* $n_i$ cases. Again, we see that the properties of the TC particles are reasonably well constrained, but not those of the TH particles. It should be noted that results are obtained from a single spectrum observed with near-nadir conditions and thus the effect of the tropospheric haze is small. Scattering in this layer will become more prominent at higher angles and it is possible that a limb-darkening analysis would provide greater constraint on the haze particle characteristics. However, such a study is beyond the scope of these IRTF/SpeX data, which only measured Uranus along its central meridian.

Also plotted in the top right panel of Figs. 3 and 4 are the required imaginary refractive indices necessary to produce single-scattering albedoes consistent with those assumed by T2013 for the TC in their empirical model. We can see that for the TC we require the imaginary part of the refractive index to increase markedly for wavelengths greater than 1 μm. This is necessary to lower the single-scattering albedo sufficiently to reduce the reflectivity of the peaks at $\lambda > 1$ μm relative to those at $\lambda < 1$ μm. This adjustment to the single-scattering albedo is very similar to that empirically arrived at by Irwin, Teanby and Davis (2010) and later by Tice et al. (2013). However, in this case we have achieved this result with a self-consistent model that is more physically plausible. For the $n_i = 0.001$ case we estimate the mean particle size in the Tropospheric Cloud (TC) to be 0.89 ± 0.04 μm, compared with the *a priori* of 1.0 ± 0.05 μm, while in the Tropospheric Haze (TH) we derive values of 0.178 ± 0.013 μm, compared with the a priori of 0.1 ± 0.05 μm (similar sizes were obtained with other a priori values of $n_i$). Hence, we see we are rather more sensitive to the size of the TH particles than we are to the size of the TC particles. However, we find we are not very sensitive to the variance of the TC and TH size distributions, which did not vary significantly from their *a priori* values of 0.05. The self-consistent retrieval model is found to fit the observed IRTF/SpeX spectrum slightly better than the emprical T2013 approach with $\chi^2/n=1.00$, compared with $\chi^2/n=1.04$. Since there were 442 points in the spectrum to be fitted this represents a change of $\Delta\chi^2 = 18$, which is significant. The retrieved cloud profiles, shown in Fig 2., can be seen to be very similar for the retrieved self-consistent and assumed empirical scattering parameter approaches. When using the empirical scattering parameter approach with the Spitzer temperature-abundance profile (shown in Fig. 6) we find the TC must reside at slightly lower pressures, as expected, since the deep methane abundance is higher, but we achieved a significantly poorer fit ($\chi^2/n=1.24$).

With this method we can compare our derived refractive indices with the published refractive index spectra of candidate condensate materials to see if some candidate particles might be more consistent than others. We shall return to this point later in Section 5, but at this point we explored the extent to which our fits depended on the assumed value of the real refractive index, $n_r$, at our reference wavelength of 1.4 μm. We performed retrievals using different assumed real refractive index values at 1.4 μm (for both haze and cloud particles) of 1.2, 1.4, 1.7, 2.0 and 2.4, with the same imaginary refractive index of 0.001. We found that the closeness of fit did depend on $n_r$ with fitted values of $\chi^2/n$ of 0.96, 1.00, 1.11, 1.52 and 1.37 respectively. Hence, particles with higher values of $n_r$ seem less able to reproduce the peak radiances at lower wavelengths and thus our analysis favours solutions with lower values of $n_r$.

## 3. Consistency of IRTF/SpeX observations with HST/STIS cloud model

A major drawback of the Irwin et al. (2009-2012a) and T2013 approach was that the methane profile was fixed to be the same at all latitudes, with a deep abundance of 1.6%, after Baines et al. (1995). The main gaseous absorber in the near-IR spectrum of Uranus is methane and to use these features to infer cloud top positions a volume mixing ratio profile must be assumed. Although we know that methane condenses in the troposphere, at wavelengths greater than 1 μm and at the lower resolutions once available there was no way to discriminate between cloud top height and methane abundance. Thus in the absence of any better information it was assumed that the same methane profile could be used at all latitudes, allowing cloud heights to be retrieved. While discrimination between methane and cloud top height is difficult at wavelengths > 1 μm, it becomes less difficult at shorter wavelengths, where methane absorption becomes less pronounced and the collision-induced absorption bands of $H_2$-$H_2$ become more prominent. Karkoschka and Tomasko (2009) (henceforth KT2009). analysed HST/STIS observations of Uranus measured in 2002 in the wavelength range 0.3 – 1 μm. Analysing the absorptions around the second overtone band of $H_2$-$H_2$ collision-induced absorption around 825 nm, KT2009 were able to show that the tropospheric mole fraction of methane at equatorial latitudes has values of ~4%, decreasing to < 2% at polar latitudes. The IRTF/SpeX observations analysed by T2013 (and reanalysed here) also include this 825 nm region and T2013 were able to confirm this enrichment of methane at equatorial latitudes. At longer wavelengths, improved spectroscopic line parameter data allowed Irwin et al. (2012b) to reanalyse Gemini/NIFS H-band observations at their native spectral resolution of R = 5370 to determine that a latitudinally varying methane abundance profile was also most consistent with these observations. Irwin et al. (2012b) found similar enhancements of methane at equatorial latitudes.

In addition to the methane-abundance/cloud-height degeneracy problem, a second puzzle for Uranus cloud studies has been discrepancies between the temperature/pressure profile previously used for Uranus cloud retrievals and the expected position of clouds from thermochemical considerations. The Voyager-2 radio-occultation observations of Uranus reported by Lindal et al. (1987) indicated a sudden change of density (presumed to be associated with the condensation level of methane) at about 1.2 bar, which coincides with the approximate level of observed discrete clouds, thought to be methane ice. Unfortunately, the temperature/pressure profile of the reference 'D' profile derived by Lindal et al. (1987) (and used in the Baines et al., 1995 model) is inconsistent with the thermal properties of methane ice, which in such a profile would condense at significantly higher pressures than 1.2 bars, especially for cases where the deep abundance of methane was as high as 4% as estimated by KT2009.

SFK2011 reanalysed the Voyager-2 radio-occulation profiles of Lindal et al. (1987). They determined that self-consistent temperature, pressure, and abundance profiles could be constructed with a deep methane abundance of 4% and methane cloud condensation level of ~1.2 bar if the He volume mixing ratio were reduced from 0.15 to 0.116, near the edge of the uncertainty range quoted by Conrath et al. (1987), and a small abundance of Ne (0.04%) were added. SFK2011 then went on to reanalyse the 2002 HST/STIS 0.3-1.0 μm observations of KT2009 and found that their 'F1' solution to the Voyager 2 radio-occultation data provided the best consistency with the observed spectra when used in conjunction with a vertical cloud model having five different particle types and layers with empirically determined scattering properties, building upon the model of KT2009. The 'F1' profile of SFK2011 is compared with the Baines et al. (1995) profile used by T2013 and with the Spitzer profile of Orton et al. (2014) in Fig. 6.

The different cloud layers and scattering properties of the SFK2011 cloud scheme are outlined in full in Table 2, and are based on empirical deductions and approximations, similar in flavour, but different from, the assumptions made by T2013. For example, for their stratospheric haze the real part of the refractive index was set to 1.4 everywhere and an empirical function used for the imaginary refractive index spectrum. For other particle types, an empirical single-scattering albedo spectrum was imposed, together with an assumed phase function. SFK2011 applied this model to the 0.3 – 1 μm HST/STIS range, but Table 2 also specifies how these parameters were extended to the 1-2 μm range (Fig. 9 of SFK2011). As can be seen the assumed scattering properties are empirical, but given that we do not have good information on the actual particle properties in Uranus' atmosphere, such assumptions are as valid as those used by T2013 and were found to be consistent with the HST/STIS and previous Voyager-2 observations.

To assess how consistent the 'F1' temperature/pressure/methane profile and cloud parameterisation scheme of SFK2011 might be with our IRTF/SpeX observations from 0.8 – 1.8 μm, we added a new parameterisation scheme to NEMESIS to split the atmosphere into layers that coincided with the edges of the 5 different cloud layers. The base pressures of the 5 cloud and haze layers were set and in the case of the three tropospheric clouds the upper pressures scaled from the base pressures with factors of 0.98, 0.93 and 0.93 for the Lower, Middle and Upper Tropospheric clouds respectively. These three clouds were each represented by a single layer in our model. The two haze layers were set to lie between 0.9 and 0.1 bar for the Tropospheric Haze and between 0.1 and 0.01 bar for the Stratospheric haze. The hazes were subdivided into 5 layers equally spaced in log pressure. The clear atmosphere between the tropospheric clouds was split into 4 layers, with an additional 4 layers placed below the Lower Tropospheric Cloud, extending to a deepest pressure of 12.4 bars and 4 layers placed above the top of the Statospheric Haze extending to lowest pressure of 0.00003 bar. Except for the thin clouds and the edges of the haze layers, which were set to specific pressure levels, all intermediate layers were equally spaced in log pressure. In all, this representation had 38 layers. Rayleigh scattering was again included.

Using this new model, the fitted opacities and pressures reported in Table 2 of SFK2011 were used to reproduce the synthetic spectra reported in their Fig. 7c. We found we were able to fully reproduce their 0.3-1.0 μm synthetic spectra with very good accuracy and also tested that our model reproduced the scattering efficiencies, single scattering albedos, backscatter phase function and backscatter efficiency shown in their Fig. 9. Having established that we were accurately reproducing the synthetic spectra of SFK2011 we then tested to see how well this model might be consistent with our IRTF/SpeX observations.

For comparison with the T2013 model, we again chose the IRTF/SpeX spectrum closest to the equator at 1.3°S, shown in Fig. 1. Figure 7 shows the agreement between our observed spectrum and the synthetic calculation using purely the SFK2011 cloud model and scattering properties. It can be seen that the SFK2011 simulates the spectrum from 0.8 – 1.1 μm very well, as we might expect, but it departs significantly from the IRTF/SpeX spectrum from 1.1 to 1.8 μm, with the radiance in the reflectance peaks and in the methane absorption bands being too high. Figure 8 shows the effect on the modelled spectrum of removing each of the five clouds in turn, where we can see that, as expected, the peak reflectivities are modulated mainly by the tropospheric clouds, while the hazes control the reflectivity in the methane absorption bands. Figure 8 shows that scattering from the upper tropospheric haze is mainly responsible for the reflection at 1.7 μm and is clearly too high. The scattering properties of this haze were modelled by SFK2011 as having a wavelength-independent cross-section and empirical single-scattering albedo and phase function, but these assumptions are clearly not valid in the 1 – 2 μm range. Similarly, we can see that the reflection in the peaks at wavelengths greater than 1.2 μm needs to be less, just as was found by Irwin, Teanby and Davis (2010) and Tice et al. (2013). This suggests that the tropospheric cloud reflectivities need to be lower at these wavelengths, either due to lower single-scattering albedo, lower cross-section, lower backscatter efficiency or a combination of all three.

Although not consistent without modification with the IRTF/SpeX observations, the SFK2011 model has significant merit in that it has a methane cloud roughly where we might expect one to be from thermochemical equilibrium for the increased deep tropospheric abundance of methane at the equator (4%) determined by KT2009. Hence, we decided to test this approach further, using their 'F1' profile in conjuction with the same 5-cloud model used by SFK2011, but replacing the scattering properties of some of the clouds with self-consistent scattering properties determined with NEMESIS using our new scheme.

**4. Reanalysis of IRTF/SpeX observations using 'F1' vertical profile and 5-cloud SFK2011 model.**

Using the SFK2011 5-cloud-layer scheme and their 'F1' temperature/abundance profile we attempted to see how well we might be able to fit the IRTF/SpeX observations. Since we found that we had little sensitivity to the pressure level of the Lower Tropospheric Cloud (LTC), we fixed its location to a base pressure of 5 bars. In this model the upper tropospheric cloud (UTC) is presumed to be the methane cloud and hence we fixed its position to the methane condensation level for the 'F1' profile of 1.23 bar. We also left the Tropospheric Haze (TH) and Stratospheric Haze (SH) between their prescribed pressure limits since previous analyses of these data have indicated that we have limited sensitivity to the vertical position of the haze, other than requiring it to reside at pressures less than ~1 bar. We did, however, allow the base pressure of the Middle Tropospheric Cloud (MTC) to vary. We left the optical properties of the Lower Tropospheric Cloud (LTC), Upper Tropospheric Cloud (UTC) and Stratospheric Haze (SH) to those specified by SFK2011, extrapolating as they suggested to the 1 – 2 μm range, but retrieved the scattering properties of the Middle Tropospheric Cloud (MTC) and Tropospheric Haze (TH) using our self-consistent refractive index retrieval method, using the same set-up as was used to determine the optical properties of the Tropospheric Cloud (TC) and Tropospheric Haze (TH) of the 2-cloud T2013 model. The optical depths of all the clouds were allowed to vary.

This model was then fitted to the near-centre-of-disc IRTF/SpeX observations at 1.3° S and the resulting fit is shown in Fig. 9, in direct comparison with the fit obtained with the simpler 2-cloud self-consistent retrieval shown earlier in Fig. 1. Similarly, Fig. 10 compares the retrieved cloud profiles (in units of optical depth/bar at 1.6 μm). The slight reduction with height of the optical depth/bar of the hazes (which is constant with height in the SFK2011 model) is due to slight differences in the way we model the aerosols in NEMESIS in terms of their specific density (particles/gram) and fractional scale height. However, these differences are not significant here where we basically have two vertically extended haze layers. We find the SFK2011 model gives a reasonably good fit to the measured spectrum, but that the fit is slightly worse than 2-cloud model ($\Delta\chi^2 = 70$, using the number of spectral points

n=442), especially in the peaks at λ < 1 μm, where the model predicts too little reflection compared with the T2013 scheme. Figure 8 shows that at these wavelengths we are sensitive to reflection from the LTC as well as the MTC and UTC, which in the SFK2011 scheme has similar scattering properties across the whole range, in contrast with the tropospheric cloud of Tice et al. (2013) which is significantly more scattering at short wavelengths than at longer wavelengths. Hence, we replaced the scattering properties of the LTC with particles having scattering properties consistent with those recommended by T2013 for the TC. We assumed a standard Gamma particle-size distribution with mean radius 1.2 μm and variance 0.05 and set the real part of the refractive index to 1.4 at a wavelength of 0.9 μm. We then empirically adjusted the imaginary refractive index spectra, computing the real part with a Kramers-Kronig calculation until we arrived at particles that had the same single-scattering albedo spectra as the T2013 Tropospheric Cloud, but self-consistent cross-section and phase-function spectra. The modified LTC parameters are listed in Table 3. Using these updated LTC scattering properties we re-ran our retrieval to derive the revised fit also shown in Fig. 9, which can be seen to fit very well at all wavelengths and fits slightly better than the 2-cloud model ($\Delta\chi^2 = 6.2$). In this retrieval we also revised the complex refractive index spectrum of the UTC to be that of methane ice (Martonchik and Orton, 1994), although we found that this change had very little effect on the calculated spectrum as the refractive indices were not greatly different from those assumed by SFK2011 and the fitted opacity of this layer was low. The revised cloud structure of the modified-LTC model is shown in Fig. 10, while the revised complex refractive index spectra of the TH and MTC (assuming an *a priori* imaginary refractive index $n_i = 0.001$) are shown in Fig. 11 (and listed in Table 4). These refractive index spectra can be seen to be almost identical to those derived for the two-cloud T2013 model, shown in Fig. 3. They were also practically identical with the refractive indices retrieved from the unmodified-LTC 5-cloud model (not shown). We also tested the sensitivity to the *a priori* values of $n_i$ and determined very similar dependencies to the simpler 2-cloud model shown earlier in Figs. 4 and 5.

Thus, by fitting the optical properties of the MTC and TH, and resetting the properties of the LTC we have derived a cloud profile that is consistent with the general structure of the SFK2011 model, but which also provides a spectrum that is consistent with IRTF/SpeX observations over the full range from 0.8 to 1.8 μm. With some refinement of the scattering properties in the 0.3 to 0.8 μm range the fit could easily be extended to the cover the HST/STIS spectral range (0.3 to 1.0 μm) also. However, we do not have access to these data and thus such an extension is beyond the scope of this paper.

As just mentioned, the retrieved imaginary refractive index spectra of the MTC and TH (shown in Fig. 11 and listed in Table 4) are very similar to those previously retrieved with the 2-cloud self-consistent model for the TC and TH (Fig. 3 and Table 1). Again, we find that the lower tropospheric clouds must have increased imaginary index of refraction at wavelengths $\lambda > 1$ μm in order to lower the single scattering albedo and thus suppress of the reflectance of the $\lambda > 1$ μm peaks relative to the $\lambda < 1$ μm peaks. The retrieved particle sizes are also similar with the mean radius of the particles in the Middle Tropospheric Cloud (MTC) being found to be 0.86±0.04 μm, while in the Tropospheric Haze we derive a mean particle size of 0.13±0.02 μm. Again, we found we were insensitive to the variances of these size distributions. Comparing the retrieved errors with the *a priori* errors, we find that again only the mean size of the particles in the TH are well constrained. This is better summarised in Table 5, where we compare the *a priori* and retrieved values (together with errors) for all non-imaginary-refractive-index parameters and include an improvement factor, which indicates how well the value of a parameter has been constrained by the retrieval. Here, it can be seen that the pressure of the MTC has not varied far from that determined by SFK2011 and lies in the region previously retrieved for Uranus' main cloud deck with this methane abundance profile.

The effect of the retrieved refractive index spectra on the actual scattering properties of the MTC and TH can be seen in Figs. 12 and 13, where we have plotted the retrieved extinction cross-section and single-scattering albedo spectra of the Middle Tropospheric Cloud (MTC) and Tropospheric Haze (TH) together with the combined Henyey-Greenstein scattering parameters that have been fitted to the Mie-calculated phase functions for the a priori $n_i$ = 0.001 case. Figures 12 and 13 also show the cloud scattering properties of Tropospheric particles assumed by SFK2011 for reference.

## 5. Discussion

The retrieved complex refractive index spectra of the particulates should help us in identifying their composition. Unfortunately, the literature contains relatively little information on the complex refractive index spectra of different candidate Uranus condensates in the near-IR range. What data that are available relate to pure, fresh condensates, and not the photochemically altered particles that are likely to exist in Uranus' atmosphere. In Fig. 14 we have compared the retrieved refractive index spectra of our Tropospheric Cloud (TC), using our best-fitting 2-cloud self-consistent model with published data for methane ice (Martonchik and Orton, 1994), ammonia ice (Martonchik, Orton and Appleby, 1984; Howett et al., 2007), solid $NH_4SH$ (Howett et al., 2007) and water droplets (Hale and Querry, 1973). As can be seen, little can be discerned from the variation in wavelength of the derived real component of the refractive index spectrum. For the imaginary part, most candidate condensates have increasing absorption at longer wavelengths, but not to the extent apparently required, suggesting there is an additional unknown source of absorption existing for the Uranus particulates. Possible candidates for the optically thickest scattering cloud at ~2 bars are $NH_3$, $H_2S$ and perhaps some component of $PH_3$ ice. However, we can see from Fig. 14 that the scattering properties of pure ammonia ice are clearly inconsistent with those required for Uranus main scattering cloud at the ~2 bar level and the scattering properties of $H_2S$ and $PH_3$ ice are unknown. Furthermore the high value of the real refractive index for $NH_4SH$ would suggest that this particle is not suitable either since we saw earlier that our method indicates that particles with real refractive index in the range 1.2-1.4 provide the best fit to the observed spectrum. Of course, there is no certainty that the particles in the main cloud deck are pure condensates at all. Photochemically produced haze particles may settle down from the stratosphere and upper troposphere and coat the fresh condensates, in a process akin to riming, and in the process mask their pure scattering properties. Using an Equilibrium Cloud Condensation Model, and assuming generally accepted deep enrichment factors of O/H = 100 × Solar, N/H = Solar, S/H = 11 × Solar and C/H = 40 × Solar (Irwin, 2009) it is found that $NH_4SH$ should condense at a pressure of ~13 bars, well below the observable level, leaving an $H_2S$ cloud to condense at ~5 bars. Of course, if the N/S ratio is closer to unity it might be that more of the $H_2S$ is combined with $NH_3$ in a deep $NH_4SH$ cloud, leaving a small remnant of $H_2S$ to condense at the lower pressures determined here. It should also be noted that if N/S > 1, then all the $H_2S$ will be combined in a lower $NH_4SH$ cloud, leaving $NH_3$ free to condense at lower pressures, although we have seen that in its pure form $NH_3$ ice particles are too reflective at all wavelengths to be consistent with the required scattering properties. This view is also inconsistent with ground-based microwave observations (de Pater and Massie, 1985), which finds that the abundances of both water and ammonia are substantially sub-solar (by factors of ~100) down to pressures of ~50 bars, but that $H_2S$ is much more abundant (10 – 35 × solar). To solve the identification of Uranus' main cloud clearly requires laboratory measurements of the complex refractive index spectrum of $H_2S$ and $PH_3$, although if the particles are heavily coated with photochemical products drizzling down from above, they are likely not pure anyway and determining their bulk composition spectroscopically may not be possible.

The latitudinal distribution of methane in Uranus' upper troposphere was found by Karkoschka and Tomasko (2009) to vary significantly with latitude, from abundances of ~4% at equatorial latitudes to < 2% near the poles. This distribution has since been confirmed by Sromovsky et al. (2011), Irwin et al. (2012b) and Tice et al. (2013). The distribution bears a remarkable similarity with ground-based microwave obsrvations of Uranus with the VLA (de Pater et al. 1991; Hofstadter and Butler 2003) which find that the deep abundance of $NH_3$ at pressures < 50 bars decreases by almost an order of magnitude from equatorial latitudes towards the poles. Such a variation is consistent with a large Hadley Cell in Uranus' atmosphere with air rising at the equator and descending at the poles, or with an atmosphere that is convectively overturning at low latitudes, but becomes convectively suppressed at latitudes polewards of 45° N,S. The fact that methane has such a similar latitudinal variation suggests that the abundances of ammonia and methane are likely linked. Such a view seems inconsistent with the circulation pattern suggested by SFK2011 (their Fig. 18), which shows a similar looking Hadley cell at pressures less than 3 bars, but which reverses at deeper pressures.

Finally, we noted earlier that our approach of modelling the scattering characteristics of ice particles with Mie scattering, which assumes spherical particles may represent a cause of systematic error, since the phase functions of spherical particles contain features, such as the 'rainbow' and 'glory', that are absent from the phase functions of a distribution of randomly orientated non-spherical particles. We have partly ameliorated this potential drawback by approximating the Mie-scattering phase functions with best-fitting combined Henyey-Greenstein phase functions, but systematic errors remain possible. We are currently in the process of updating our model to use pre-tabulated scattering properties calculated for a range of non-spherical particles by Dobovik et al. (2006), which should correct for such potential errors (although adding another degree of freedom to the model – the particle shape). However, this development is still in progress and hence beyond the scope of this paper.

## 6. Conclusions

The retrieval mechanism developed here represents a significant improvement over previous modelling studies of near-IR Uranus spectra. By retrieving the imaginary refractive index spectra of the clouds and constructing the real part of the refractive index through a Kramers-Kronig analysis, we are able to determine self-consistent extinction cross-section, single-scattering and phase-function spectra for the clouds and hazes in Uranus' atmosphere over the range 0.8 to 1.8 μm. We find that we require particles in the main cloud deck at ~2 bars to be significantly more absorbing at wavelengths $\lambda > 1.0$ μm than they are at shorter wavelengths $\lambda < 1.0$ μm, which provides a potential constraint on the identity of the main particle type. We find that the scattering properties required are inconsistent with pure ammonia ice and we thus recommend that laboratory studies be conducted to establish if $H_2S$ or $PH_3$, which might also potentially condense at these pressures, have such scattering properties, which would then potentially identify them as being the main cloud condensate in Uranus' atmosphere.

In addition to developing the new retrieval scheme we compared the 2-cloud-layer model of Tice et al. (2013), using the methane profile of Baines et al. (1995) with the 5-cloud-layer model of Sromovsky, Fry and Tomasko (2011), using their reanalysed 'F1' Voyager 2 radio-occultation profile. We note here that we also fitted the spectra with the 2-cloud scheme using the 'F1' profile and found a reasonable fit, but poorer than that obtained when using the Baines et al. (1995) profile and worse also than when using the Spitzer profile (Orton et al. 2014). We find that both 2-cloud and 5-cloud models reproduce well the observed centre-of-disc IRTF/Spex spectrum of Uranus, recorded in 2009. However, it can be seen in Fig. 9 that a modified version of the SFK2011 model does fit the observations significantly better at shorter wavelengths and the 'F1' profile has the added advantage of being based on a temperature/abundance profile that is simultaneously consistent with the enhanced equatorial abundances of methane determined by Karkoschka and Tomasko (2009) and leads to methane condensation at 1.2 bars, where the Voyager 2 radio-occultation profile noted a distinct step in the refractivity gradient. We should note, however, that the 'F1' profile of SFK2011 requires a $He/H_2$ ratio that lowers the tropopause temperatures, and indeed the entire temperature profile above the $CH_4$ condensation level, to values too cold to be consistent with Spitzer IRS observations. We find that we retrieve similar refractive index spectra for the main cloud at ~2 bars using both approaches. However, this analysis does not lead to a very clear indication of which profile is most consistent with the IRTF/SpeX observations, although the 5-cloud model does seem to improve slightly the fit at wavelengths less than 1 μm.

**Acknowledgements**

We are grateful to the United Kingdom Science and Technology Facilities Council for funding this research. Leigh Fletcher was supported by a Royal Society Research Fellowship at the University of Oxford.

**Tables**

| | Tropospheric Cloud | | | | Tropospheric Haze | | | |
|---|---|---|---|---|---|---|---|---|
| $\lambda$ | $n_r$ | $n_i$ | $\chi$ | $\varpi$ | $n_r$ | $n_i$ | $\chi$ | $\varpi$ |
| 0.7 | 1.40014 | 0.00101 | 0.99498 | 0.98547 | 1.39989 | 0.00100 | 13.35856 | 0.99096 |
| 0.8 | 1.39874 | 0.00096 | 1.14211 | 0.98950 | 1.39990 | 0.00099 | 9.25143 | 0.98939 |
| 0.9 | 1.39702 | 0.00068 | 1.26472 | 0.99405 | 1.39990 | 0.00099 | 6.59613 | 0.98760 |
| 1.0 | 1.39448 | 0.00080 | 1.33341 | 0.99411 | 1.39993 | 0.00103 | 4.82549 | 0.98481 |
| 1.1 | 1.38987 | 0.00102 | 1.34484 | 0.99338 | 1.39996 | 0.00102 | 3.59473 | 0.98224 |
| 1.2 | 1.38033 | 0.00418 | 1.28960 | 0.97522 | 1.39997 | 0.00101 | 2.71645 | 0.97939 |
| 1.3 | 1.38607 | 0.05149 | 1.13529 | 0.77422 | 1.39998 | 0.00101 | 2.07951 | 0.97570 |
| 1.4 | 1.40000 | 0.01206 | 1.20073 | 0.93726 | 1.40000 | 0.00099 | 1.61016 | 0.97211 |
| 1.5 | 1.38728 | 0.03687 | 1.05920 | 0.83024 | 1.40000 | 0.00097 | 1.26102 | 0.96826 |
| 1.6 | 1.39696 | 0.06189 | 1.00000 | 0.74939 | 1.39999 | 0.00098 | 1.00000 | 0.96281 |
| 1.7 | 1.43592 | 0.00467 | 1.11423 | 0.97672 | 1.39997 | 0.00102 | 0.80312 | 0.95540 |
| 1.8 | 1.42657 | 0.00118 | 1.01875 | 0.99390 | 1.39998 | 0.00103 | 0.65161 | 0.94815 |
| 1.9 | 1.42104 | 0.00100 | 0.93042 | 0.99472 | 1.40002 | 0.00101 | 0.53326 | 0.94201 |

Table 1. Derived scattering properties of the Tropospheric Cloud and Tropospheric Haze using 2-cloud Tice et al. (2013) model, with the *a priori* imaginary refarctive index, $n_i$, set to 0.001±50% at all wavelengths. The imaginary part of the refractive index spectrum ($n_i$) is retrieved by the model and a Kramers-Kronig analysis used to construct the real part ($n_r$). The complex refractive index spectrum is then used in a Mie-scattering calculation to constuct the extinction efficiency ($\chi$, normalised to 1.6 μm) and single-scattering albedo ($\varpi$) spectra. The errors on these fitted and derived quantities is best understood by inspection of Figures 4 and 5, which show the retrieved quantities obtained for different *a priori* values of $n_i$.

| Cloud | $p_{Base}$ (bar) | $p_{Top}$ (bar) | Cross section | Single scattering albedo | Phase function |
|---|---|---|---|---|---|
| Stratospheric Haze | 0.1 | 0.01 bar | $n_r$=1.4, $n_i(\lambda)$ = 0.0055exp[(350-$\lambda$)/100] ($\lambda$ in nm), extrapolated from 1-2 μm. Scattering properties computed with Mie scattering, assuming standard Gamma distribution of sizes with $r_0$ = 0.1 μm and variance=0.3. | | |
| Tropospheric Haze | 0.9 | 0.1 | Constant | $\varpi(\lambda)$=1−1/[2+exp[($\lambda$-290)/37]] ($\lambda$ in nm), extrapolated from 1-2 μm. | Henyey Greenstein: g1=0.7, g2=-0.3, f($\lambda$)=0.94-0.47sin4[(1000-l)/445] ($\lambda$ in nm). f($\lambda$) fixed to f(1.0) at all wavelengths from 1-2 μm. |
| Upper Tropospheric Cloud (UTC) | 1.23 | 1.14 (1.28 × 0.93) | N=1.4+0i at all wavelengths. Scattering properties computed with Mie scattering, assuming standard Gamma distribution of sizes with $r_0$ = 1.2 μm and variance=0.1 | | |
| Middle Tropospheric Cloud (MTC) | 1.68 | 1.56 (1.68 × 0.93) | As Tropospheric Haze | | |
| Lower Tropospheric Cloud (LTC) | 5 | 4.9 (5 × 0.98) | As Tropospheric Haze | | |

Table 2. Five-cloud layering scheme of Sromovsky, Fry and Karkoschka (2011) and associated cloud scattering properties.

| λ | $n_r$ | $n_i$ | χ | ϖ |
|---|---|---|---|---|
| 0.7 | 1.36675 | 0.00001 | 0.88527 | 0.99978 |
| 0.8 | 1.38547 | 0.00001 | 0.91996 | 0.99981 |
| 0.9 | 1.39103 | 0.00001 | 1.00000 | 0.99984 |
| 1 | 1.39029 | 0.00001 | 1.11674 | 0.99987 |
| 1.1 | 1.39802 | 0.01875 | 1.17604 | 0.84218 |
| 1.2 | 1.39552 | 0.0375 | 1.20952 | 0.76656 |
| 1.3 | 1.38551 | 0.05625 | 1.20510 | 0.72023 |
| 1.4 | 1.38071 | 0.075 | 1.17185 | 0.68199 |
| 1.5 | 1.37764 | 0.075 | 1.16517 | 0.69163 |
| 1.6 | 1.37547 | 0.075 | 1.14720 | 0.69793 |
| 1.7 | 1.37385 | 0.075 | 1.12095 | 0.70173 |
| 1.8 | 1.37258 | 0.075 | 1.08888 | 0.70362 |
| 1.9 | 1.37157 | 0.075 | 1.05294 | 0.70403 |

Table 3. Empically determined scattering properties of Lower Tropospheric Cloud (LTC) of the SFK2011 scheme adjusted to match the single-scattering albedo spectrum of tropospheric scatterers determined by Tice et al. (2013). A standard gamma size distribution was assumed with mean radius 1.2 μm and variance 0.05. Again, the extinction efficiency (here normalised to 0.9 μm) and single-scattering albedo spectra were determined from the complex refractive index spectrum using Mie theory.

|   | Middle Tropospheric Cloud | | | | Tropospheric Haze | | | |
|---|---|---|---|---|---|---|---|---|
| $\lambda$ | $n_r$ | $n_i$ | $\chi$ | $\varpi$ | $n_r$ | $n_i$ | $\chi$ | $\varpi$ |
| 0.7 | 1.39713 | 0.00099 | 1.08248 | 0.98682 | 1.39990 | 0.00100 | 17.31323 | 0.98629 |
| 0.8 | 1.39606 | 0.00097 | 1.24206 | 0.99024 | 1.39991 | 0.00099 | 11.51517 | 0.98306 |
| 0.9 | 1.39478 | 0.00089 | 1.35722 | 0.99276 | 1.39991 | 0.00098 | 7.87520 | 0.97908 |
| 1.0 | 1.39261 | 0.00075 | 1.40963 | 0.99476 | 1.39993 | 0.00101 | 5.51755 | 0.97327 |
| 1.1 | 1.38851 | 0.00097 | 1.40061 | 0.99391 | 1.39995 | 0.00102 | 3.95039 | 0.96676 |
| 1.2 | 1.38180 | 0.00628 | 1.32371 | 0.96418 | 1.39997 | 0.00101 | 2.88757 | 0.95978 |
| 1.3 | 1.38813 | 0.04229 | 1.18941 | 0.80838 | 1.39998 | 0.00101 | 2.15559 | 0.95117 |
| 1.4 | 1.40000 | 0.00764 | 1.23239 | 0.95983 | 1.40000 | 0.00100 | 1.63888 | 0.94242 |
| 1.5 | 1.38767 | 0.02202 | 1.09363 | 0.89127 | 1.40000 | 0.00098 | 1.26803 | 0.93294 |
| 1.6 | 1.38985 | 0.05468 | 1.00000 | 0.76820 | 1.39999 | 0.00100 | 1.00000 | 0.91996 |
| 1.7 | 1.42617 | 0.00405 | 1.08911 | 0.97949 | 1.39999 | 0.00102 | 0.80171 | 0.90488 |
| 1.8 | 1.41844 | 0.00112 | 0.98855 | 0.99411 | 1.40000 | 0.00103 | 0.65106 | 0.88990 |
| 1.9 | 1.41397 | 0.00100 | 0.89883 | 0.99459 | 1.40003 | 0.00101 | 0.53427 | 0.87613 |

Table 4. Derived scattering properties of Middle Tropospheric Cloud (MTC) and Tropospheric Haze (TH) using the modified SFK2011 model with modifed Lower Tropospheric Cloud (LTC) properties, with the *a priori* imaginary refarctive index, $n_i$, set to 0.001±50% at all wavelengths. The imaginary part of the refractive index spectrum ($n_i$) is retrieved by the model and a Kramers-Kronig analysis used to construct the real part ($n_r$). The errors on these fitted and derived quantities is best understood by inspection of Figures 4 and 5, which show the retrieved quantities obtained for different *a priori* values of $n_i$.

| Variable | $x_a$ | $e_a$ | $e_a/x_a$ | $x_n$ | $e_n$ | $e_n/x_n$ | Improve(%) |
|---|---|---|---|---|---|---|---|
| $P_{LTC}$ (bar) | 5.0 | fixed | - | 5 | fixed | - | - |
| $\tau_{LTC}$ | 3.45 | 3.45 | 1.00 | 7.20 | 0.997 | 0.139 | 86.1 |
| $P_{MTC}$ (bar) | 1.68 | 0.05 | 0.0298 | 1.78 | 0.025 | 0.014 | 53.1 |
| $\tau_{MTC}$ | 1.28 | 1.28 | 1.00 | 1.49 | 0.077 | 0.0521 | 94.8 |
| $P_{UTC}$ (bar) | 1.23 | fixed | - | 1.23 | fixed | - | - |
| $\tau_{UTC}$ | 0.322 | 0.322 | 1.00 | 0.087 | 0.0195 | 0.224 | 77.6 |
| $\tau_{TH}$ | 0.01 | 0.01 | 1.00 | 0.0019 | 0.000129 | 0.068 | 93.2 |
| $\tau_{SH}$ | 0.0012 | 0.0012 | 1.00 | 0.0003 | 0.000083 | 0.274 | 72.6 |
| $r_{MTC}$ | 1.00 | 0.05 | 0.05 | 0.86 | 0.037 | 0.0433 | 13.3 |
| $b_{MTC}$ | 0.05 | 0.01 | 0.2 | 0.0495 | 0.0099 | 0.1994 | 0.3 |
| $r_{TH}$ | 0.1 | 0.05 | 0.5 | 0.1309 | 0.0205 | 0.1565 | 68.7 |
| $b_{TH}$ | 0.05 | 0.01 | 0.2 | 0.0507 | 0.0101 | 0.1994 | 0.3 |

Table 5. Retrieved non-imaginary-refractive-index parameters for the 5-cloud SFK2011 model with modified LTC applied to the IRTF/SpeX centre-of-disc Uranus spectra. Here $x_a$, $e_a$ are the *a priori* values and errors, $x_n$, $e_n$ are the retrieved values and errors. These parameters are held as log(values) by NEMESIS and hence the improvement factor is defined as $100 \times \left(1 - (e_n/x_n)/(e_a/x_a)\right)$ and indicates how far the retrieved error limits differ from the *a priori*. Values with a high improvement factor are well constrained by the retrieval.

**Figures**

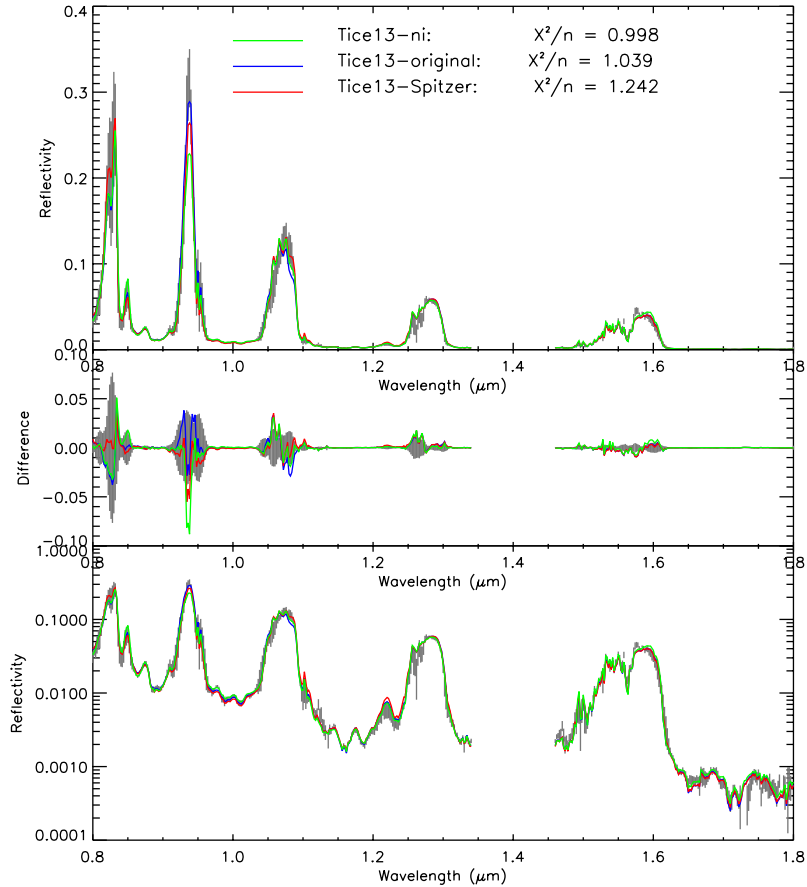

Figure 1. Measured IRTF/SpeX spectrum at 1.3ºS (thick grey line indicates mesurement and random error) and fit to it with the 2-cloud model of T2013 using three different methods: 1) self-consistent scattering property retrieval (green), 2) original T2013 model with empirically determined scattering properties (blue) and 3) original T2013 model with empirically determined scattering properties, but using the Sptizer temperature-abundance profile (red). The top panel shows the reflectivity spectrum (I/F) in linear space, with the differences shown in the middle panel. The bottom panel shows the reflectivity space in log space to accentuate the strong methane-absorbing regions. The $\chi^2/n$ of the fits are shown in the top panel. The spectrum between 1.34 and 1.46 μm is not accurately measured by IRTF due to telluric water absorption and so has been omitted here.

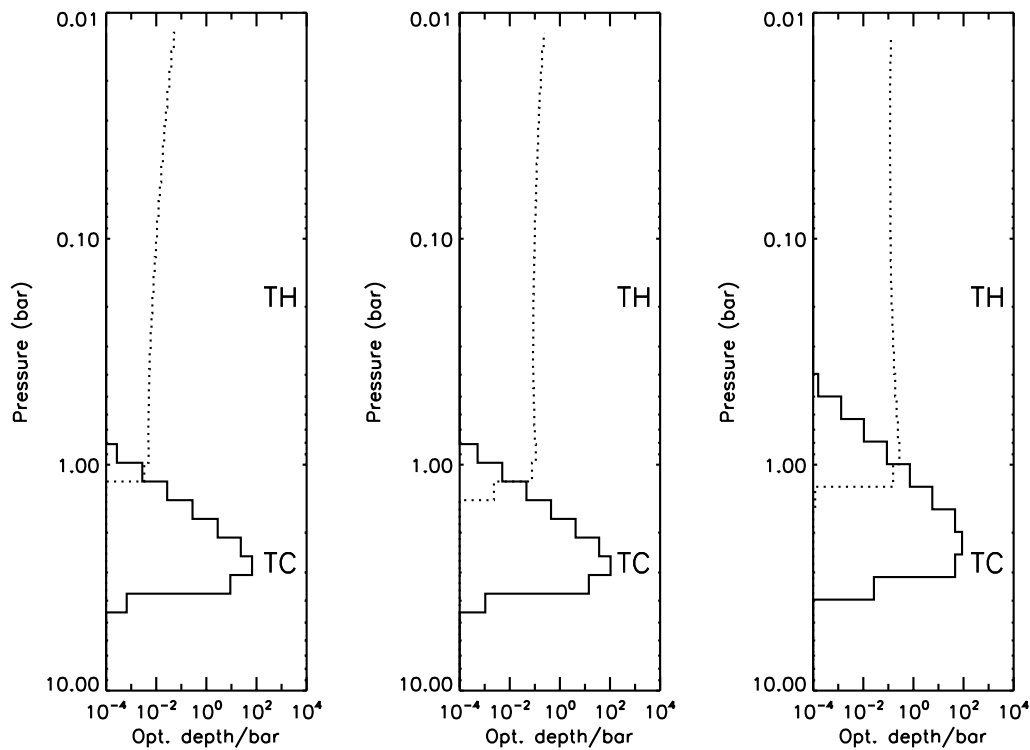

Figure 2. Fitted cloud density profiles (in units of optical depth/bar at 1.6 μm) for the 2-cloud T2013 model using: 1) self-consistent scattering particles retrieval and T2013 atmospheric model (left panel), 2) T2013's empirically adjusted scattering parameters and the T2013 atmospheric model (middle panel), and 3) T2013's empirically adjusted scattering parameters and the Spitzer atmospheric model (right panel). As can be seen very similar cloud structres are inferred with the self-consistent retrieval approach and the original T2013 method. For the Spitzer temperature-abundance profile, which has a higher deep methane abundance, the TC moves to slightly lower pressures as expected.

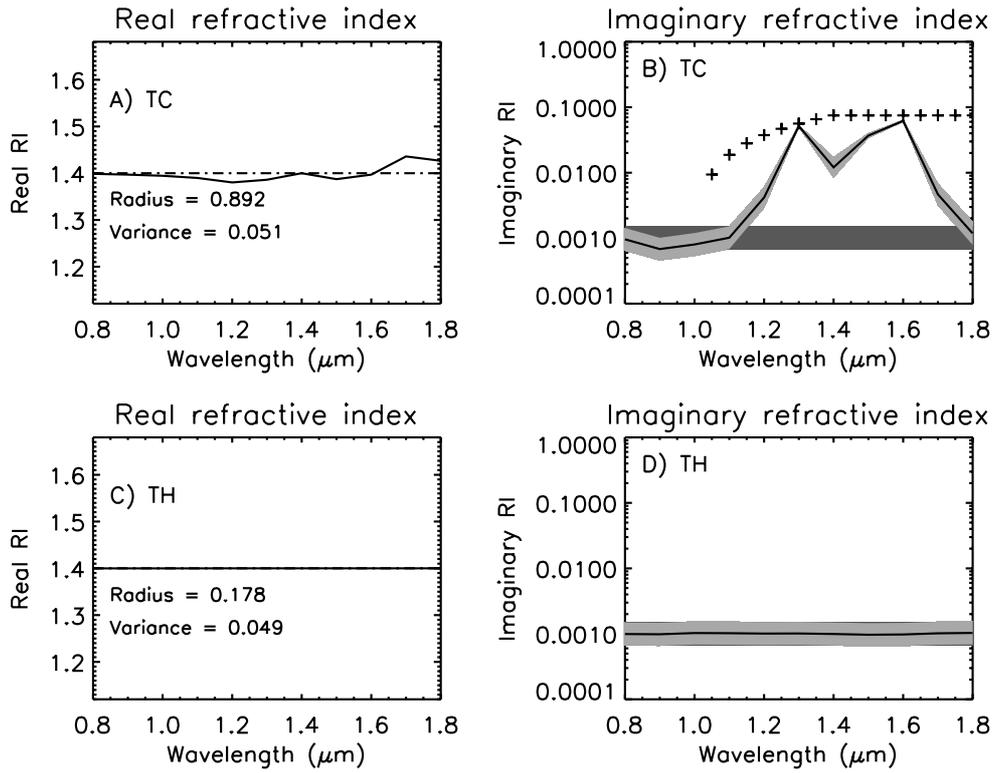

Figure 3. Fitted complex refractive indices of the Tropospheric Cloud (TC) and Tropospheric Haze (TH) using the 2-cloud model with our self-consistent scattering parameter retrieval, with *a priori* imaginary refractive index for both particle types set to 0.0010±0.0005 (i.e. 50%). The *a priori* imaginary refractive index spectrum and errors are shown as the dark grey region, while the fitted spectrum and errors are shown as the light grey region and superposed solid lines. The *a priori* real refractive index spectra are shown by the dot-dash lines and the inferred Kramers-Kronig spectrum by the solid line. The cross symbols show the imaginary refractive indices necessary to reproduce the single-scattering albedo spectra inferred for the TC by Tice et al. (2013). Comparing the size of the error bars we can see that the imaginary refractive index of the TC particles is well constrained in the absorption peaks at 1.3 ands 1.6 μm, but is less well contrained elsewhere. The imaginary refractive index of the TH particles is not well constrained using this single observed spectrum.

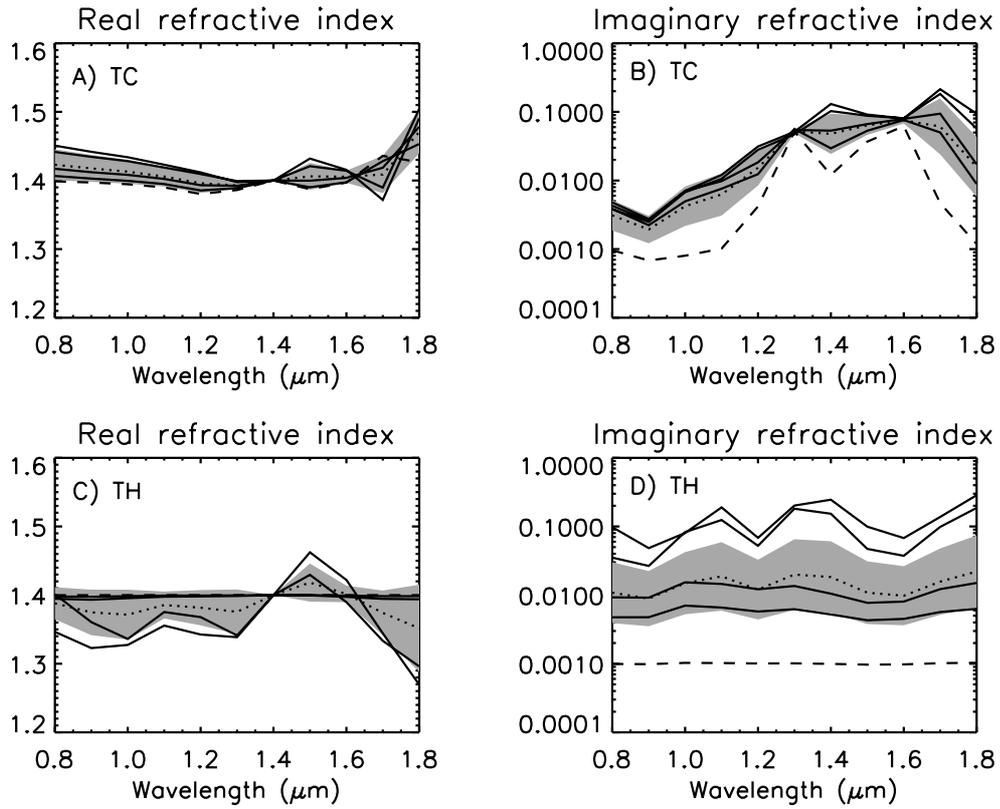

Figure 4. Fitted complex refractive indices of the Tropospheric Cloud (TC) and Tropospheric Haze (TH) using the 2-cloud model with self-consistent scattering parameter retrieval, with *a priori* imaginary refractive index for both particle types set to 0.001, 0.005, 0.01, 0.05 and 0.1 (all ±50%). The solid lines are the individual cases, except the case where $n_i$=0.001 for which a dashed line is used. The mean of all five cases are the dotted lines with standard deviation indicated by the grey region. The same variation of imaginary refractive index with wavelength is found in all five cases for the TC particles, but there is much less constraint on the TH particles.

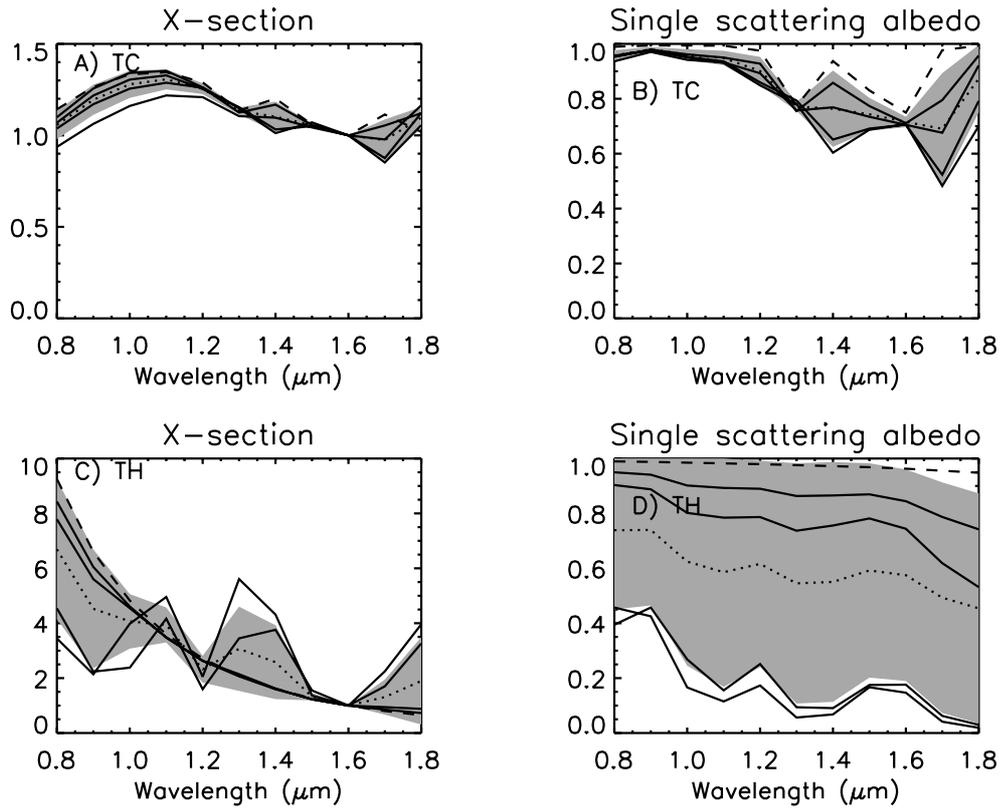

Figure 5. Fitted cross-section (normalised at 1.6 μm) and single scattering albedoes of the Tropospheric Cloud (TC) and Tropospheric Haze (TH) using the 2-cloud model with self-consistent scattering parameter retrieval, with *a priori* imaginary refractive index for both particle types set to 0.001, 0.005, 0.01, 0.05 and 0.1 (all ±50%). The line styles are identical to those used in Fig. 4. As can be seen the proporties of the TC particles are well constrained, but there is considerable degeneracy in the solutions for the TH particles.

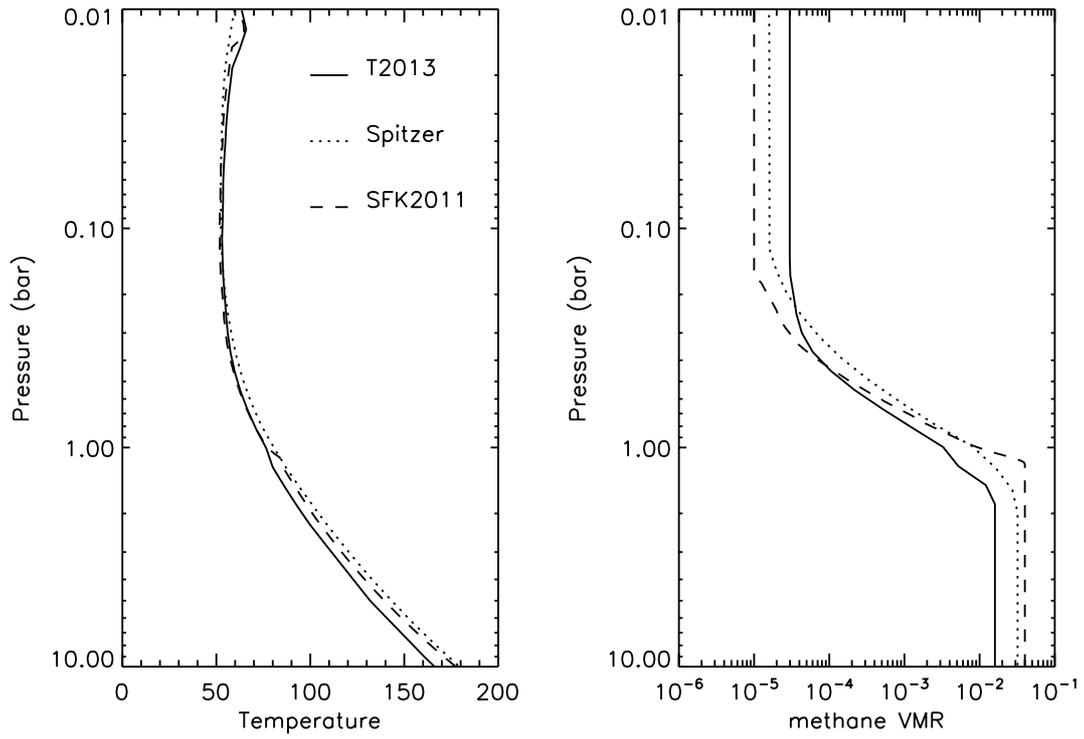

Figure 6. Comparison of the different vertical profiles of temperature and methane assumed in this study, namely the Baines et al. (1995) profile used by T2013, the Spitzer profile (Orton et al., 2014), and the 'F1' profile of SFK2011.

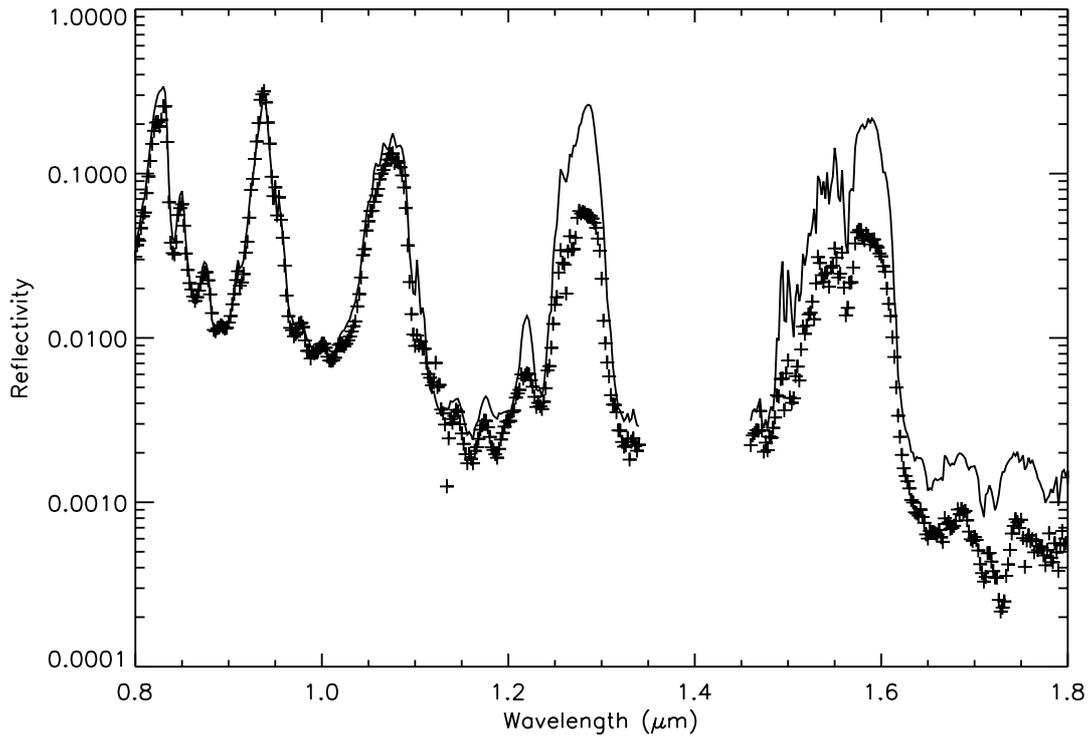

Figure 7. Measured Uranus spectrum observed by IRTF/SpeX at 1.3°S (cross symbols) together with synthetic spectrum calculated with the published SFK2011 cloud model, without any modification. The synthetic spectrum at wavelengths less than 1 μm, which the model was designed to fit, shows a good correspondence with the IRTF/SpeX spectrum (although the reflectance in the 0.8 and 1.15 μm peaks is a little low), but the agreement becomes increasingly poor at longer wavelengths.

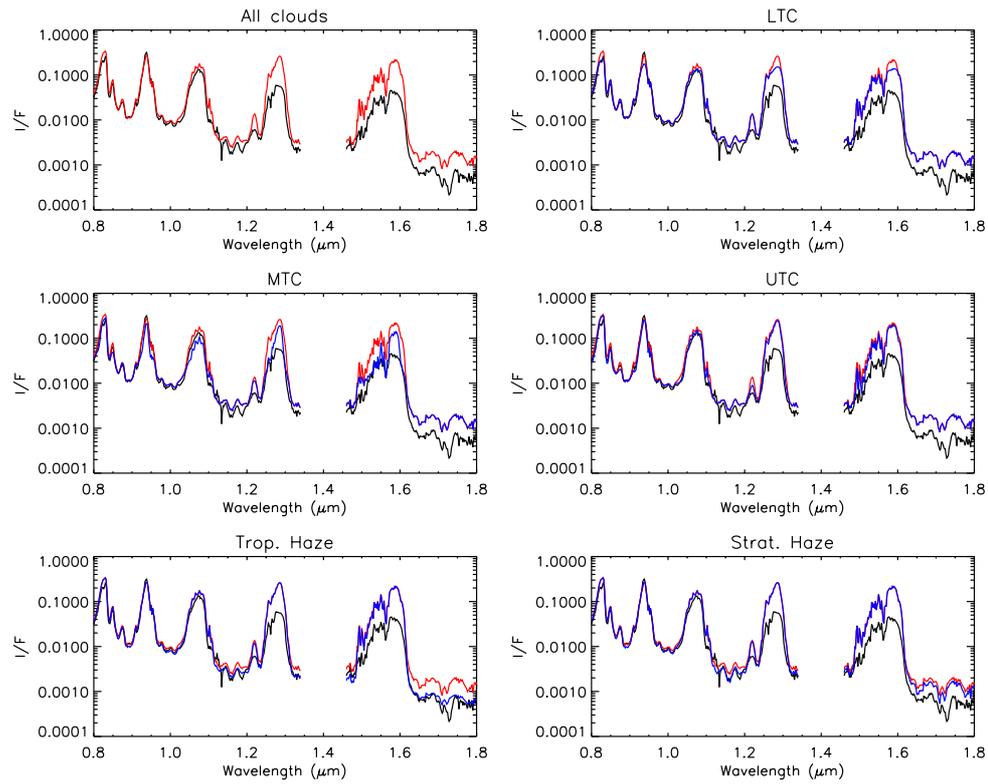

Figure 8. Effect of removing in turn the contribution to the synthetic spectrum of each cloud layer of the SFK2011 cloud model. In each plot the measured IRTF/SpeX spectrum at 1.3°S is indicated by the black line and the spectrum calculated from the standard SFK2011 model is shown in red. The spectrum calculated with one of the cloud layers removed is shown by the blue lines in the separate panels.

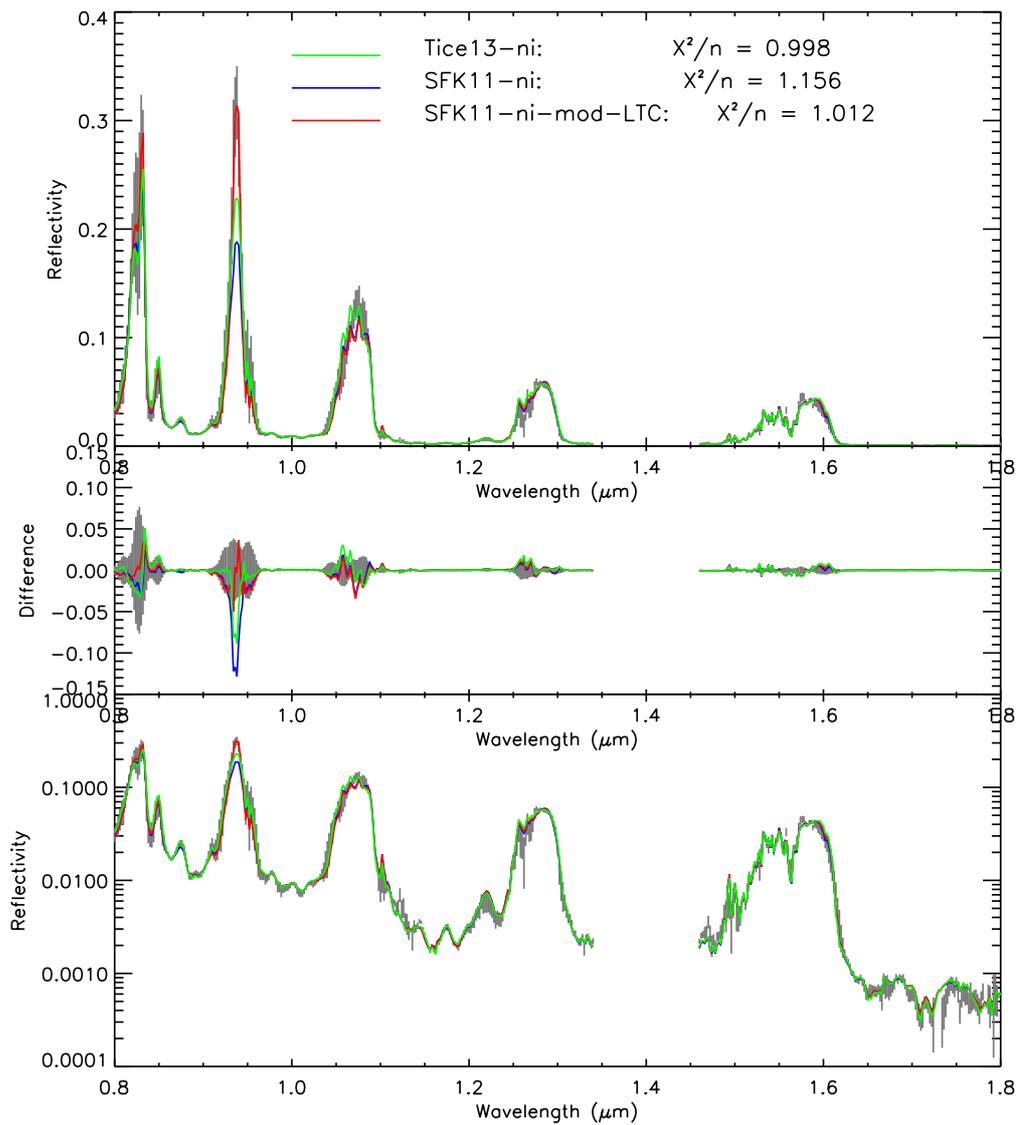

Figure 9. Measured IRTF/SpeX spectrum at 1.3ºS (thick grey line indicates mesurement and random error) and fit to it with with self-consistent retrieval model using 1) a 2-cloud model of T2013 (green), 2) the first modification of the cloud model of SFK2011 (blue), and the final modification of this model (red) with modified LTC properties. Again the top panel shows the reflectivity spectrum (I/F) in linear space with the differences shown in the middle panel. The bottom panel shows the reflectivity space in log space. The $\chi^2/n$ of the fits are shown in the top panel. The spectrum between 1.34 and 1.46 μm is not accurately measured by IRTF due to telluric water absorption and so has been omitted here.

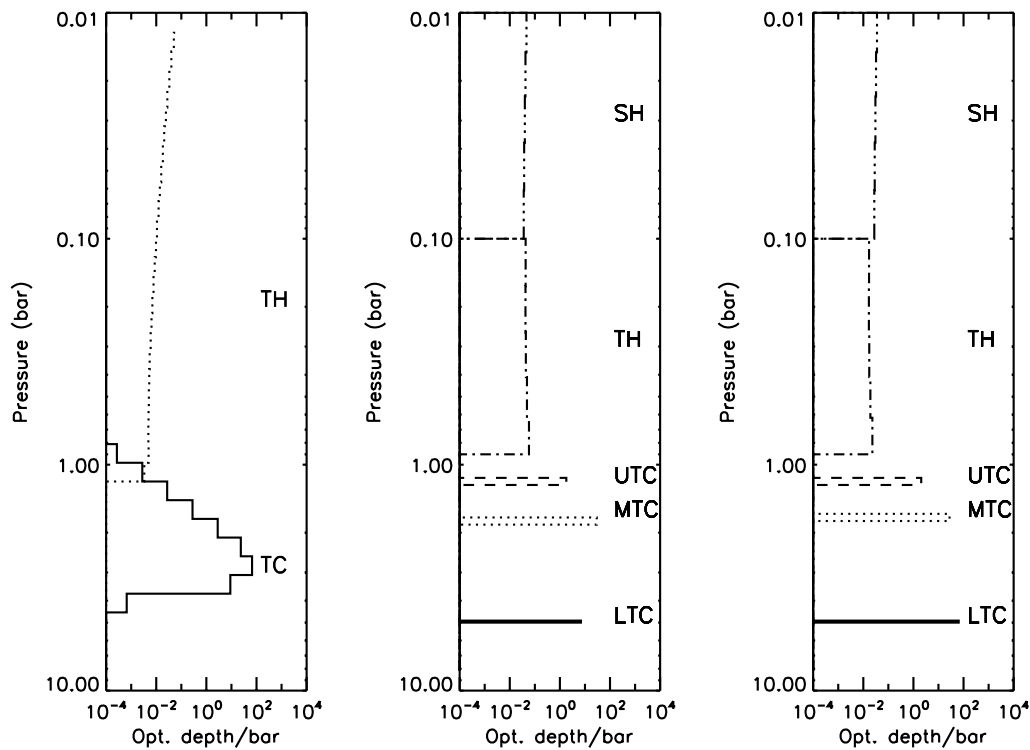

Figure 10. Fitted cloud density profiles (in units of optical depth/bar at 1.6 μm) for the: 1) 2-cloud self-consistent model (left panel), 2) the modified SKF2011 model using their LTC properties (middle), and the modified SKF2011 model using revised LTC properties consistent with the TC of T2013 (right panel). Opacities are here referred to their values at 1.6 μm, not to the normalising wavelength of 0.9 μm used by SFK2011.

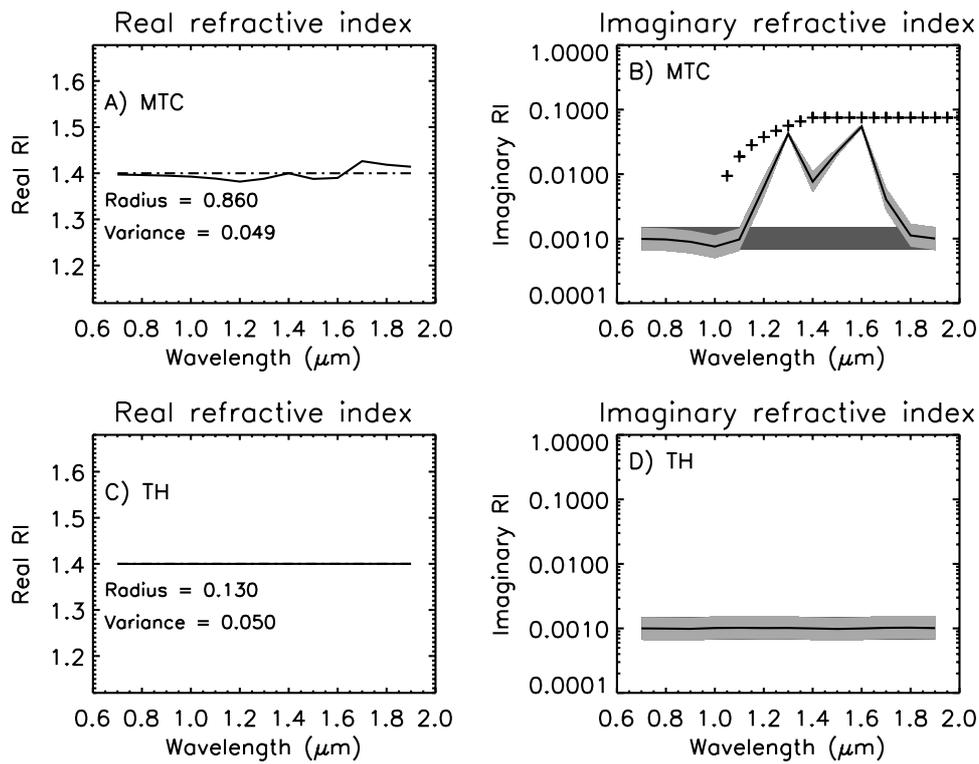

Figure 11. Fitted spectral properties of the Middle Tropospheric Cloud (MTC) and Tropospheric Haze (TH) using the modified SFK2011 5-cloud model, with revised spectral properties of the Lower Tropospheric Cloud (LTC). The linestyles are those defined in Fig. 3 and the cross symbols again show the imaginary refractive indices necessary to reproduce the single-scattering albedo spectra inferred for the TC by Tice et al. (2013). It was these properties that were assumed for the LTC here.

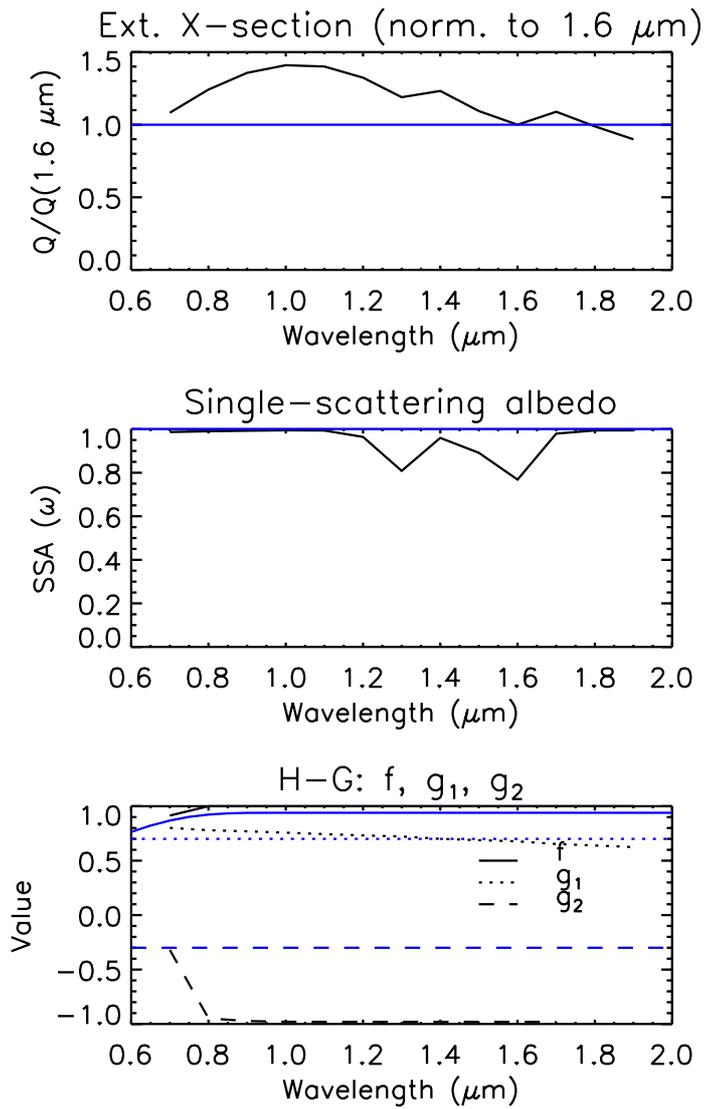

Figure 12. Retrieved scattering properties of the Middle Tropospheric Cloud (MTC) using the 5-cloud SFK2011 model, with modified LTC properties, with a priori imaginary refractive index set to 0.001±50%. The corresponding scattering properties assumed by SFK2011 for tropospheric particles are marked in blue.

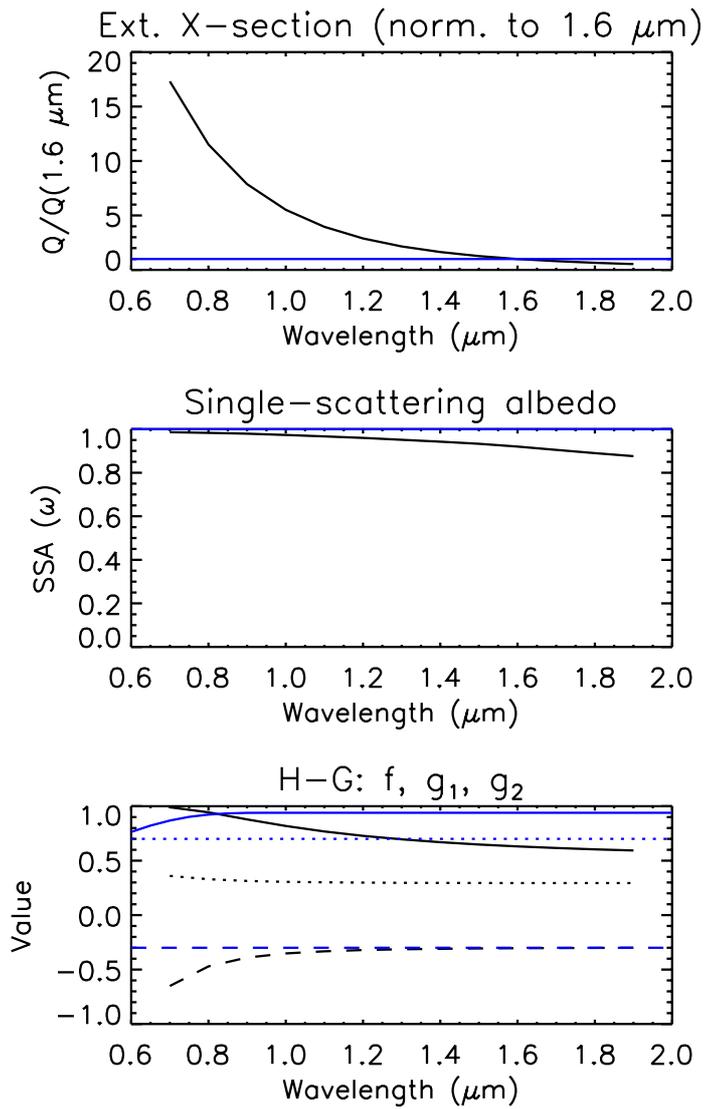

Figure 13. Retrieved scattering properties of the Tropospheric Haze (TH) using the modifed 5-cloud SFK2011 model, with revised LTC properties, with a priori imaginary refractive index set to 0.001±50%. The scattering properties assumed by SFK2011 for tropospheric particles are marked in blue.

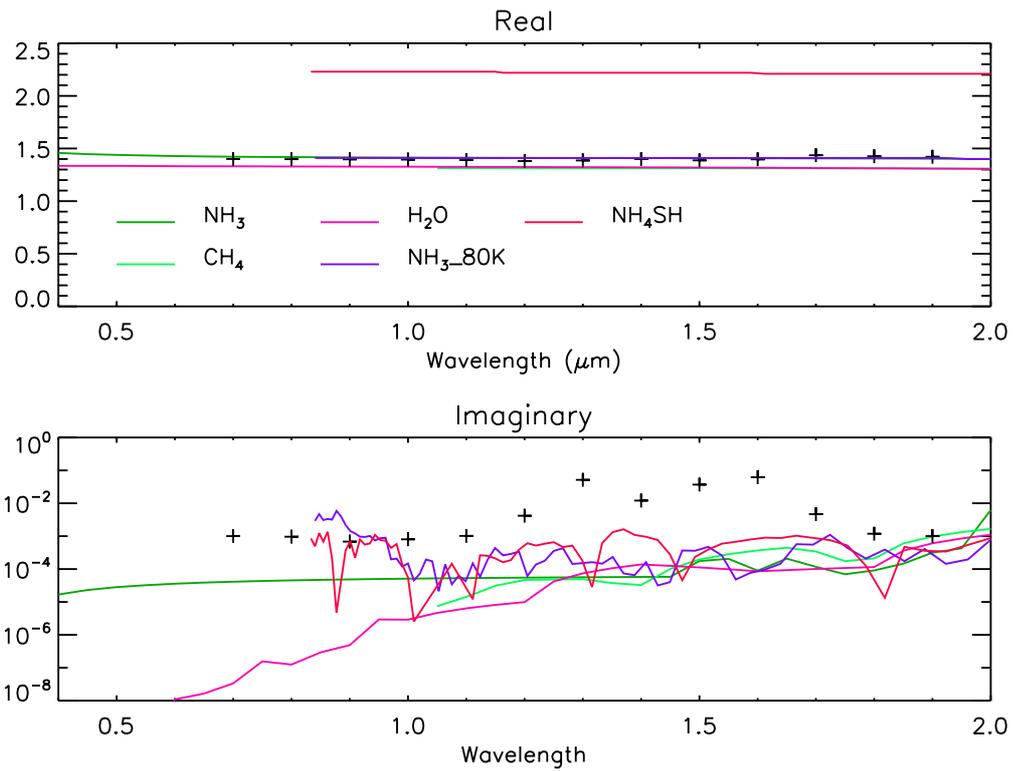

Figure 14. Comparison of retrieved tropospheric cloud (TC) refractive indices (for self-consistent 2-cloud model) with published literature referred to in the text. There are two sources of $NH_3$ data: '$NH_3$' is from Martonchik, Orton and Appleby (1984), while '$NH_3\_80K$' is from Howett et al. (2007). The retrieved TC refractive indices are indicated by the cross symbols.